\documentclass[a4paper,11pt]{article}
\usepackage[dvips]{graphicx}
\usepackage{amssymb}
\usepackage{amsmath}

\begin{document}

\title{Inverse and Dynamical Supersymmetry Breaking in $S^1\times R^3$}
\author{V.K.Oikonomou\thanks{
voiko@physics.auth.gr}\\
Dept. of Theoretical Physics Aristotle University of Thessaloniki,\\
Thessaloniki 541 24 Greece\\
and\\
T.E.I. Serres} \maketitle

\begin{abstract}
In this paper we study the influence of hard supersymmetry
breaking terms in a $N=1$, $d=4$ supersymmetric model, in
$S^1\times R^3$ spacetime topology. It is found that for some
interaction terms and for certain values of the couplings,
supersymmetry is unbroken for small lengths of the compact radius,
and breaks dynamically as the radius increases. Also for another
class of interaction terms, when the radius is large supersymmetry
is unbroken and breaks dynamically as the radius decreases. It is
pointed out that the two phenomena have similarities with the
theory of metastable vacua at finite temperature and with the
inverse symmetry breaking of continuous symmetries at finite
temperature (where the role of the temperature is played by the
compact dimension's radius).

\end{abstract}

\bigskip
\section*{Introduction}

Supersymmetry serves as the most promising extension of the
Standard Model. In the short future supersymmetry will be verified
experimentally through the LHC experiments. The elegance of the
theory and the simplifications that introduces is of great
importance however supersymmetry must be broken in our four
dimensional world. There exist various mechanisms of supersymmetry
breaking. In this article we are interested in dynamical breaking
of supersymmetry induced from a toroidal compact dimension.
Theories with one toroidal compact dimension resemble
(mathematically) field theories at finite temperature. It is known
that supersymmetry is spontaneously broken at finite temperature
\cite{fuji}, a fact that is closely related to the boundary
conditions that fermions and bosons obey in the ''thermal''
compact dimension. Specifically supersymmetry is spontaneously
broken due to the periodicity of bosonic degrees of freedom and
anti-periodicity of the fermionic degrees of freedom. Field
theories at finite temperature are conceptually related to field
theories with compact dimensions. Thus it is easily understood
that the boundary conditions of the fields in the compact
dimensions control the breaking of supersymmetry. In general,\
when studying supersymmetric theories in flat spacetime, the
background metric is ordinary Minkowski. Spacetime topology affect
the boundary conditions of the fields that are integrated in the
path integral.\ Given a class of metrics, several spacetime
topologies are allowed.\ Here we shall focus on a model that has
${S}^{1}{\times R}^{3}$ topology underlying the spacetime, $S^{1}$
refers to a spatial dimension.\ The specific topology is a
homogeneous topology of the flat Clifford--Klein type
\cite{ellis}. \ Non trivial topology, implies non trivial field
configurations, that enter dynamically in the action. This non
triviality enters in the action through the boundary conditions
and as is well known the boundary conditions are controlled from
the topology. The effective potential is a strong order parameter
indicating when supersymmetry is broken. The appearance in the
effective potential of vacuum terms which have different
coefficients for fermions and bosons lead to the fact that the
effective potential of the theory has no longer its minimum at
zero and thus, supersymmetry is spontaneously broken.\ This quite
general phenomenon can only be avoided in field theories with
compact spatial dimensions, if in some way these vacuum terms are
cancelled \cite{love}. Indeed the combination of the allowed
boundary conditions, as we show later, can save spontaneous
supersymmetry breaking at finite volume. Particularly this is due
to the fact that we can have periodic fermions and anti-periodic
bosons. This cannot be avoided at finite temperature due to the
restricted boundary conditions as we already mentioned, except in
cases where fermions have a complex chemical potential
\cite{bartolome} as in pure $SU(N)$ Yang Mills theories with
adjoint fermions at finite temperature \cite{kogan}, namely,
\begin{equation}\label{adjointfermions}
\psi (\beta )=-e^{\frac{i2\pi k}{N}} \psi (0).
\end{equation}
Normally a question arises while following the above
considerations. Why should someone care in avoiding spontaneous
supersymmetry breaking at finite volume? This is because when
supersymmetry spontaneously breaks (like at finite temperature)
then supersymmetry ceases to be a controllable symmetry of the
theory, since it is always broken and not dynamically, but by
definition. One would like to have control on the way that
supersymmetry breaks (especially in our case that the model
spacetime we work is 4 dimensional and supersymmetry must be a
symmetry and be broken dynamically. This can be avoided in
theories with higher dimensions where supersymmetry can hold in
the higher dimensional space and break in our world).

In this paper we shall study a 4 dimensional N=1 supersymmetric
model at one loop, in topology ${S}^{1}{\times R}^{3}$. We shall
find the allowed  field configurations that are determined from
the topology and construct in a correct way the Lagrangian. The
calculation of the effective potential follows, through which we
shall find when supersymmetry breaks and when does not. After that
we shall add in the Lagrangian non holomorphic and hard
supersymmetry breaking terms. This terms break supersymmetry
hardly. However for some values of the couplings and of the
masses, interesting phenomena occur. Particularly it is found that
in some cases when the volume of the compact space is small
supersymmetry is not broken and when the compact radius exceeds a
critical value, supersymmetry breaks dynamically. We shall give a
cosmological implication of this case which resembles first order
cosmological phase transitions in the early universe. Also other
terms have a curious but worthy of mentioning effect. For these
terms another strange phenomenon occurs for specific values of the
couplings and masses. In detail for large values of the
circumference of the compact dimension, supersymmetry is unbroken
in contrast to the previous case and breaks after a critical
length, as the compact dimension magnitude decreases. Although
this is not so interesting from a phenomenological point of view
(in four spacetime dimensions), it worths mentioning (and may have
application in the physics of extra dimensions). There exist
similar works in the literature but with the difference that the
symmetry under study is not supersymmetry but a global symmetry
(continuous or discrete). At finite temperature in some cases
broken $O(N_f)\times O(N_{\psi})$ symmetries become restored at
high temperatures. Also unbroken symmetries at small temperatures
may break at high temperatures (a phenomenon known as inverse
symmetry breaking). The last may have cosmological implications.
In our case the same occurs but with supersymmetry in place of the
symmetry and for a compact dimension playing the temperature's
role (also note that roughly the high temperature limit is closely
related to the small length limit through the transformation
$T\rightarrow \frac{1}{L}$. Actually through the last
transformation we can relate the two limits where this is
possible). One interesting feature of this resemblance is the
similarity of the terms in the lagrangian that trigger these
phenomena in both cases. Also these terms appear in the new
inflationary models and help in the procedure of reheating the
universe after inflation. We shall present and describe everything
in detail in the forthcoming sections. Also let us mention that
our calculations will be in 1-loop level and within the
perturbative limits with $mL\leq 1$, where $m$ is the largest mass
scale in the theory and $L$ is the circumference of the compact
dimension. Also within the four dimensional setup we use,
renormalizability of masses and couplings is ensured when $mL\leq
1$.

In section 1 we review the mathematical setup needed for field
theories with non trivial topology. The resemblance with extra
dimensional theories is pointed out. In section 2 we describe the
N=1 supersymmetric model we shall use and calculate the effective
potential in the case of the compact dimension having infinite
length and after that at finite volume. In section 3 we add
several supersymmetry breaking terms and study in detail their
effect on the vacuum energy of the model. In section 4 we review
the continuous symmetry restoration, symmetry non-restoration and
inverse symmetry breaking at finite temperature and point out the
resemblance of these with our case (a resemblance that stems from
the interactions of the scalar sector). In section 5 we present a
cosmological application of one of our results and in section 6 a
short discussion with the conclusions follow.

\section{Non Trivial Topology, Twisted Fields and Supersymmetry}

The existence of non trivial field configurations (in terms of
boundary conditions) due to non trivial topology (twisted fields),
was first pointed out by Isham \cite{isham} and then adopted by
other authors \cite{gongcharov,ford,spalluci}.\ In the spacetime
of our case, the topological properties of $S^{1}{\times R}^{3}$
are classified by the first Stieffel class $H^{1}(S^{1}{\times
R}^{3},Z_{\widetilde{2}})$ which is isomorphic to the singular
(simplicial) cohomology group ${H}_{1}({S} ^{1}{\times
R}^{3}{,Z}_{2})$ because of the triviality of the
${Z}_{\widetilde{2}}$ sheaf. It is known that
$H^{1}{(S}^{1}{\times R}^{3}{,Z}_{\widetilde{2}}{)=Z}_{2}$
classifies the twisting of a bundle. Specifically, it describes
and classifies the orientability of a bundle globally. In our
case, the classification group is ${Z}_{2}$ and, we have two
locally equivalent bundles, which are however different globally
{\it{i.e.}}~cylinder-like and moebius strip-like. The mathematical
lying behind, is to find the sections that correspond to these two
bundles, classified completely by $Z_{2}$ \cite{isham}. The
sections we shall consider are real scalar fields and Majorana
spinor fields which carry a topological number called moebiosity
(twist), which distinguishes between twisted and untwisted fields.
The twisted fields obey anti-periodic boundary conditions, while
untwisted fields periodic in the compactified dimension. Usually
(inspired by field theory at finite temperature) one takes scalar
fields to obey periodic and fermion fields anti-periodic boundary
conditions, disregarding all other configurations that may arise
from non trivial topology. We shall consider all these
configurations. Let $\varphi _{u}$, $\varphi_{t}$ and $\psi _{t}$,
$\psi _{u}$ denote the untwisted and twisted scalar and twisted
and untwisted spinor fields respectively. The boundary conditions
in the ${S}^{1}$ dimension are,
\begin{equation}\label{bc1}
\varphi_{u}(x,0)=\varphi _{u}(x,L),
\end{equation}
and
\begin{equation}\label{bc2}
\varphi _{t}(x,0)=-\varphi _{t}(x,L),
\end{equation}
for scalar fields and
\begin{equation}\label{bc1}
\psi _{u}(x,0)=\psi _{u}(x,L),
\end{equation}
and
\begin{equation}\label{bc2}
\psi_{t}(x,0)=-\psi _{t}(x,L),
\end{equation}
for fermion fields, where $x$ stands for the remaining two spatial
and one time dimension which are not affected by the boundary
conditions. Spinors (both Dirac and Majorana), still remain
Grassmann quantities. We assign the untwisted fields twist
${h}_{0}$  (the trivial element of ${Z}_{2})$ and the twisted
fields twist $h_{1}$  (the non
trivial element of ${Z}_{2}$).\ Recall that $h_{0}+h_{0}=h_{0}$ ($0+0=0$), $%
h_{1}+h_{1}=h_{0}$ ($1+1=0$), $h_{1}+h_{0}=h_{1}$ ($1+0=1$). We
require the Lagrangian to be scalar under $Z_{2}$ thus to have
${h}_{0}$ moebiosity. The
topological charges flowing at the interaction vertices must sum to ${h}_{0}$ under ${H}^{1}{(S}^{1}{\times R}^{3}{,Z%
}_{\widetilde{2}}{)}$. For supersymmetric models, supersymmetry
transformations impose some restrictions on the twist assignments
of the superfield component fields \cite{gongcharov}.

Now which fields can acquire vacuum expectation value? Grassmann
fields cannot acquire vacuum expectation value (vev) since we
require the vacuum value to be a scalar representation of the
Lorentz group. Thus, the question is focused on the two scalars.
The twisted scalar cannot acquire non zero vev \cite{ford},
consequently, only untwisted scalars are allowed to develop vev's.

In the literature, twisted fields have frequently been used, for
example in the Scherk-Schwarz mechanism \cite{scherk}, where the
harmonic expansion of the fields is of the form:
\begin{equation}
\phi(x,y)=e^{imy}\sum_{n=-\infty }^{\infty
}{\phi}_{n}(x)e^{\frac{i2{\pi}ny}{L}},
\end{equation}
The ''$m$'' parameter incorporates the twist mentioned above. This
treatment is closely related to automorphic field theory
\cite{Dowker} in more than 4 dimensions (which is an alternative
to the one used by us). The Scherk-Schwarz mechanism is a well
known mechanism that generates supersymmetry breaking to our 4
dimensional world after dimensional reduction and is frequently
used for compactifications in extra dimensional models.

Concerning the automorphic field theory, let us quote here a
different approach to the above. Due to the compact dimension we
can use generic boundary conditions for bosons and fermions in the
compact
dimension. These are,%
\begin{eqnarray}
\varphi _{i}(x_{2},x_{3},\tau ,x_{1}) &=&e^{i\pi n_{1}\alpha
}\varphi
_{i}(x_{2},x_{3},\tau ,x_{1}+L) \\
~\Psi (x_{2},x_{3},\tau ,x_{1}) &=&e^{i\pi n_{1}\delta }\Psi
(x_{2},x_{3},\tau ,x_{1}+L),  \notag
\end{eqnarray}%
with, $0<\alpha ,\delta <1$, $i=1,2$, $n_{1}=1,2,3...$. The values
$\alpha= 0,1$ correspond to periodic and antiperiodic bosons
respectively while $\delta =0,1$ corresponds to periodic and
anti-periodic fermions (for details see \cite{Dowker}).

\section{Description of the Supersymmetric Model}

The model we shall present is described by the global ${N=1}$,
${d=4}$ supersymmetric Lagrangian,
\begin{equation}
\mathcal{L}=[\Phi _{1}^{+}\Phi _{1}]_{D}+[\Phi ^{+}\Phi
]_{D}+[\frac{m_{1}}{2}\Phi ^{2}+\frac{g_{1}}{6}\Phi
^{3}+\frac{m}{2}\Phi _{1}^{2}+g\Phi \Phi _{1}^{2}]_{F}+{\rm{H.c}},
\label{lagra}
\end{equation}%
where $\Phi _{1}$, $\Phi $ are chiral superfields and the
superpotential from which the interaction part of the lagrangian
arises is $[\frac{m_{1}}{2}\Phi ^{2}+\frac{g_{1}}{6}\Phi
^{3}+\frac{m}{2}\Phi _{1}^{2}+g\Phi \Phi _{1}^{2}]_{F}$. In the
above,
\begin{eqnarray}
\Phi &=&\varphi _{u}(x)+\sqrt{2}\theta \psi _{u}(x)+\theta \theta
F_{\varphi _{u}}+i\partial _{\mu }\varphi _{u}(x)\theta \sigma ^{\mu }\bar{\theta } \\
&&-\frac{i}{\sqrt{2}}\theta \theta \partial _{\mu }\psi _{u}(x)\sigma ^{\mu }%
\bar{\theta }-\frac{1}{4}\partial _{\mu }\partial ^{\mu }\varphi
_{u}^{+}(x)\theta \theta \bar{\theta }\bar{\theta }, \notag
\end{eqnarray}%
is a left chiral superfield. It contains the untwisted scalar
field components and the untwisted Weyl fermion. Although the
untwisted scalar is complex, we shall use the real components
which will be the representatives of the
sections of the trivial bundle classified by ${H}^{1}{(S}^{1}{\times R}^{3}{%
,Z}_{\widetilde{2}}{)}$. Moreover,%
\begin{eqnarray}
\Phi _{1} &=&\varphi _{t}(x)+\sqrt{2}\theta \psi _{t}(x)+\theta
\theta F_{\varphi _{t}}+i\partial _{\mu }\varphi _{t}(x)\theta
\sigma ^{\mu }\bar{\theta }
\\
&&-\frac{i}{\sqrt{2}}\theta \theta \partial _{\mu }\psi _{t}(x)\sigma ^{\mu }%
\bar{\theta }-\frac{1}{4}\partial _{\mu }\partial ^{\mu }\varphi
_{t}^{+}(x)\theta \theta \bar{\theta }\bar{\theta }, \notag
\end{eqnarray}%
is another left chiral superfield containing the twisted scalar
field and the twisted Weyl fermion. Writing down (\ref{lagra}) in
component form, we
get (writing Weyl fermions in the Majorana representation):%
\begin{align}
\mathcal{L}& =\partial _{\mu }\varphi _{u}^{+}\partial ^{\mu
}\varphi _{u}-\left \vert m_{1}\varphi _{u}+\frac{g_{1}}{2}\varphi
_{u}\varphi _{u}+g\varphi _{t}^{2}\right \vert
^{2}+i\overline{\Psi }_{t}\gamma ^{\mu }\partial _{\mu
}\Psi _{t}-\frac{1}{2}m\overline{\Psi }_{t}\Psi _{t} \label{test}\\
& -\frac{g_{1}}{4}(\overline{\Psi }_{u}\Psi _{u}-\overline{\Psi
}_{u}\gamma
_{5}\Psi _{u})\varphi _{u}-\frac{g_{1}}{4}(\overline{\Psi }_{u}\Psi _{u}+%
\overline{\Psi }_{u}\gamma _{5}\Psi _{u})\varphi _{u}^{+}+\partial
_{\mu
}\varphi _{t}^{+}\partial ^{\mu }\varphi _{t}-  \notag \\
& \left \vert m\varphi _{t}+2g\varphi _{t}\varphi _{u}\right \vert ^{2}+i%
\overline{\Psi }_{u}\gamma ^{\mu }\partial _{\mu }\Psi _{u}-\frac{1}{2}m_{1}%
\overline{\Psi }_{u}\Psi _{u}-  \notag \\
& \frac{g}{4}(\overline{\Psi }_{t}\Psi _{t}-\overline{\Psi
}_{t}\gamma
_{5}\Psi _{t})\varphi _{u}-\frac{g}{4}(\overline{\Psi }_{t}\Psi _{t}+%
\overline{\Psi }_{t}\gamma _{5}\Psi _{t})\varphi _{u}^{+}.  \notag
\end{align}%
Notice that moebiosity is conserved at all interaction vertices
{\it{i.e.}}~equals ${h}_{0}$. The moebiosity of\ $\varphi _{u}$
and $\Psi _{u}$ is ${h}_{0}$ while for $\varphi _{t}$\ and\ $\Psi
_{t}$ is ${h}_{1}$. Using the ${Z}_{2}$ cyclic group properties we
see that the Lagrangian (\ref{test}) has moebiosity ${h}_{0}$.\
The complex field $\varphi _{u}$ can be written in terms of real
components as ${\varphi }_{u}{=\chi +i\varphi
}_{u_{2}}{/}\sqrt{2}$, where $\chi ={v+(\varphi
}_{u_{1}}{)/}\sqrt{2}$ (${v}$ is the classical value).\ Thus,
$\varphi _{u_{1}}$\ and $\varphi _{u_{2}}$ are
real untwisted field configurations belonging to the trivial element of ${%
H}^{1}{(S}^{1}{\times R}^{3}{,{Z}_{\widetilde{2}})}$ and
satisfying periodic boundary conditions in the compactified
dimension.\ The twisted scalar field
can be written as ${\varphi }_{t}{=(\varphi }_{t_{1}}{+i\varphi }_{t_{2}}{)/}%
\sqrt{2}$, since, this field, being a member of the non trivial
element of ${H}^{1}{(S}^{1}{\times R}^{3}{,Z}_{\widetilde{2}}{)}$
cannot acquire a vev. Notice we gave a vev for an untwisted boson,
This is useful in order to find the minimum of the effective
potential minimizing it in terms of $v$. The masses of the two
Majorana fermion fields and the four bosonic fields at tree order are calculated to be:%
\begin{eqnarray}
m_{b_{1}}^{2} &=&m_{1}^{2}+3g_{1}m_{1}v+3g_{1}^{2}v^{2}/2 \label{mass}\\
\ m_{b_{2}}^{2} &=&m_{1}^{2}+g_{1}m_{1}v+g_{1}^{2}v^{2}/2  \notag \\
m_{t_{1}}^{2}\bigskip &=&m^{2}+4gmv+4g^{2}v^{2}+g^{2}m_{1}v/\sqrt{2}%
+g^{2}g_{1}v^{2}/4\   \notag \\
\ m_{t_{2}}^{2}
&=&m^{2}+4gmv-g^{2}m_{1}v/\sqrt{2}-g^{2}g_{1}v^{2}/4  \notag
\\
m_{f_{1}} &=&m_{1}+g_{1}v,\ m_{f_{2}}=m+2gv.  \notag
\end{eqnarray}%
In (\ref{mass}) $m_{b_{1}}$, $m_{b_{2}}$ are the masses of the
untwisted bosons (${\varphi }_{u_{1}}$ and ${\varphi }_{u_{2}}$
respectively), $m_{t_{1}}$, $m_{t_{2}}$ are the masses of the
twisted bosons (${\varphi }_{t_{1}}$ and ${\varphi }_{t_{2}}$)
and, finally, $m_{f_{1}}$, $m_{f_{2}}$ are the untwisted Majorana
fermion and twisted Majorana fermion masses respectively ($\Psi
_{u}$ and $\Psi _{t}$). The general tree level result for theories
with rigid supersymmetry in terms of chiral superfields is
satisfied (see \cite{martin}) {\it{i.e.}}~:
\begin{equation}
STr(M^{2})=\sum \limits_{j}(-1)^{2j}(2j+1)m_{j}^{2}=0.
\label{str}
\end{equation}%
Also, the following relations hold true:%
\begin{equation}
m_{b_{1}}^{2}+m_{b_{2}}^{2}=2m_{f_{1}}^{2},\
m_{t_{1}}^{2}+m_{t_{2}}^{2}=2m_{f_{2}}^{2}.\label{sp}
\end{equation}%
Since twisted scalars cannot acquire vacuum expectation value,
supersymmetry is not spontaneously broken at tree level, like in
the O' Raifeartaigh models. Indeed the auxiliary field equations,%
\begin{eqnarray}
F_{\varphi _{u}}^{+} &=&m_{1}\varphi _{u}+\frac{g_{1}}{2}\varphi
_{u}^{2}+g\varphi _{t}^{2}=0 \\
F_{\varphi _{t}}^{+} &=&m\varphi _{t}+2g\varphi _{u}\varphi
_{t}=0, \notag
\end{eqnarray}%
imply that $\varphi _{u}=0$ and $\varphi _{t}=0$ and consequently
$v=0$, thus, at tree level, no spontaneous supersymmetry breaking
occurs.

\subsection{Supersymmetric Effective Potential in $S^{1}\times R^{3}$}

We now proceed by assuming that the topology is changed to
${S}^{1}{\times R}^{3}$, while the local geometry remains
Minkowski. The metric reads:
\begin{equation}
\mathrm{d}s^{2}=\mathrm{d}t^{2}-\mathrm{d}x_{1}^{2}-\mathrm{d}x_{2}^{2}-\mathrm{d}x_{3}^{2}
\label{flat},
\end{equation}
with $-\infty <x_{2},x_{3},t<\infty $ and $0<x_{1}<L$ with the points${\ x}%
_{1}=0~$and ${x}_{1}=L$ periodically identified. The boundary
conditions for the fields are:%
\begin{align}
~\varphi _{u}(x_{1},x_{2},x_{3},t) &=~~~~~\varphi _{u}(x_{1}+L,x_{2},x_{3},t) \\
~\varphi _{t}(x_{1},x_{2},x_{3},t) &=~{}-\varphi
_{t}(x_{1}+L,x_{2},x_{3},t)
\notag \\
~\Psi _{u}(x_{1},x_{2},x_{3},t) &=~~~~~\Psi
_{u}(x_{1}+L,x_{2},x_{3},t) \notag
\\
~\Psi _{t}(x_{1},x_{2},x_{3},t) &=~~-\Psi
_{t}(x_{1}+L,x_{2},x_{3},t). \notag
\end{align}%
We Wick rotate the time direction ${t}\rightarrow {it}$\ thus
giving the background metric the Euclidean signature
\cite{spalluci}. The twisted fermions and twisted bosons will be
summed over odd Matsubara frequencies, while the untwisted
fermions and untwisted scalars will be summed over even Matsubara
frequencies \cite{dolan,coleman weinberg}. Adopting the
$\overline{DR}^{\prime }$ renormalization scheme \cite{martin} the
Euclidean effective potential
at one loop level reads:%
\begin{align}
V& =V_{0}+\frac{1}{64\pi ^{2}L}\sum \limits_{n=-\infty }^{\infty }\int \frac{%
\mathrm{d}^{3}k}{(2\pi )^{3}}{\bigg (}\ln [k^{2}+\frac{4\pi^{2} n^{2}}{L^{2}}%
+m_{b_{1}}^{2}] \\
& -2\ln [k^{2}+\frac{4\pi^{2} n^{2}}{L^{2}}+m_{f_{1}}^{2}]+\ln [k^{2}+\frac{%
4\pi^{2} n^{2}}{L^{2}}+m_{b_{2}}^{2}]  \notag \\
& -2\ln [k^{2}+\frac{\pi^{2} (2n+1)^2}{L^{2}}+m_{f_{2}}^{2}]+\ln [k^{2}+\frac{%
\pi^{2} (2n+1)^2}{L^{2}}+m_{t_{1}}^{2}]  \notag \\
& +\ln [k^{2}+\frac{\pi^{2} (2n+1)^2}{L^{2}}+m_{t_{2}}^{2}]{\bigg
).} \notag
\end{align}%
${V}_{0}~$includes the tree and the one loop corrections for infinite length,%
\begin{align}
V_{0}& =m_{1}^{2}v^{2}+g_{1}^{2}m_{1}v^{3}+\frac{g_{1}^{2}v^{4}}{4}+\frac{1}{%
64\pi ^{2}}{\bigg (}m_{b_{1}}^{4}(\ln [\frac{m_{b_{1}}^{2}}{\mu ^{2}}]-%
\frac{3}{2}) \\
& +m_{b_{2}}^{4}(\ln [\frac{m_{b_{2}}^{2}}{\mu ^{2}}]-\frac{3}{2}%
)+m_{t_{1}}^{4}(\ln [\frac{m_{t_{1}}^{2}}{\mu ^{2}}]-\frac{3}{2}%
)+m_{t_{2}}^{4}(\ln [\frac{m_{t_{2}}^{2}}{\mu ^{2}}]-\frac{3}{2})  \notag \\
& -2m_{f_{1}}^{4}(\ln [\frac{m_{f_{1}}^{2}}{\mu ^{2}}]-\frac{3}{2}%
)-2m_{f_{2}}^{4}(\ln [\frac{m_{f_{2}}^{2}}{\mu
^{2}}]-\frac{3}{2}){\bigg )}, \notag
\end{align}%
and $\mu $ is the renormalization scale, being of the order of the
largest mass \cite{tamvakis}. Furthermore, we shall assume that
$m{L\simeq 1}$ which is required for the validity of perturbation
theory \cite{guth, s weinberg}.

\subsection{How Can Supersymmetry be Broken Spontaneously in $S^1\times R^3$}

It is well known that when one considers only twisted fermions and
untwisted bosons in ${S}^{1}{\times R}^{3}$ (like in thermal field
theories), vacuum contributions $\sim
$${L}^{-4}$ do not cancel and supersymmetry is spontaneously
broken. The non-cancellation occurs because bosons and fermions
satisfy different boundary conditions. In our model the field
content is such that cancellation of vacuum contributions is being
enforced, after having included all topologically inequivalent
allowed field configurations. This situation is similar to finite
temperature calculations.\ The question if supersymmetry is broken
or not requires to check the zero modes of the vacuum state
\cite{witten}.\ It is easy to see why in conventional finite
temperature field theories and their conceptional analogues
${S}^{1}{\times R}^{3}$ topological field theories supersymmetry
is broken.\ The vacuum state, in the Wess--Zumino in
${S}^{1}{\times R}^{3}$, does contain one bosonic zero mode. In
our case this does not occur because we have equal vacuum zero
modes (twisted spinors, do not have a zero mode).\ Consequently,
in our model, we expect that supersymmetry will not be
spontaneously broken \cite{love} (for a detailed discussion we
recommend the paper of Fujikawa \cite{fuji}).

\subsection{Small $L$ Expansion of the effective potential}

The leading order contribution to the one loop effective potential
for small $L$ values is given by \cite{dolan,elizalde,kirsten}:
\begin{align}
&V=m_{1}^{2}v^{2}+g_{1}^{2}m_{1}v^{3}+\frac{g_{1}^{2}v^{4}}{4} \label{pot}\\
&{-\frac{3(2m_{f_{1}}^{4}-m_{b_{1}}^{4}-m_{b_{2}}^{4})}{4096\pi ^{4}}-\frac{%
3(2m_{f_{2}}^{4}-m_{t_{1}}^{4}-m_{t_{2}}^{4})}{256\pi ^{4}}}\notag\\
&{+\frac{%
3(2m_{f_{1}}^{4}-m_{b_{1}}^{4}-m_{b_{2}}^{4}+2m_{f_{2}}^{4}-m_{t_{1}}^{4}-m_{t_{2}}^{4})%
}{128\pi ^{2}}}\notag \\
&{+\frac{(\gamma -\ln [4\pi ])(2m_{f_{1}}^{4}-m_{b_{1}}^{4}-m_{b_{2}}^{4})}{%
1024\pi ^{4}}+\frac{(\gamma +\ln [\frac{2}{\pi }%
])(2m_{f_{2}}^{4}-m_{t_{1}}^{4}-m_{t_{2}}^{4})}{64\pi ^{4}}}\notag \\
&{+\frac{(2m_{f_{1}}^{3}-m_{b_{1}}^{3}-m_{b_{2}}^{3})}{384L\pi ^{3}}-\frac{%
(2m_{f_{1}}^{2}-m_{b_{1}}^{2}-m_{b_{2}}^{2})}{768\pi ^{2}L^{2}}+\frac{%
(2m_{f_{2}}^{2}-m_{t_{1}}^{2}-m_{t_{2}}^{2})}{384\pi ^{2}L^{2}}}\notag\\
&{+\frac{2m_{f_{1}}^{4}\ln [Lm_{f_{1}}]-m_{b_{2}}^{4}\ln
[Lm_{b_{2}}]-m_{b_{1}}^{4}\ln [Lm_{b_{1}}]}{1024\pi ^{4}}}\notag \\
&{+\frac{2m_{f_{2}}^{4}\ln [Lm_{f_{2}}]-m_{t_{2}}^{4}\ln
[Lm_{t_{2}}]-m_{t_{1}}^{4}\ln [Lm_{t_{1}}]}{64\pi ^{4}}}\notag \\
&{-\frac{(2m_{f_{1}}^{4}\ln [\frac{m_{f_{1}}^{2}}{\mu ^{2}}]-m_{b_{2}}^{4}\ln [%
\frac{m_{b_{2}}^{2}}{\mu ^{2}}]-m_{b_{1}}^{4}\ln
[\frac{m_{b_{1}}^{2}}{\mu
^{2}}])}{64\pi ^{2}}}\notag\\
&{-\frac{(2m_{f_{2}}^{4}\ln [\frac{m_{f_{2}}^{2}}{\mu ^{2}}]-m_{t_{2}}^{4}\ln [%
\frac{m_{t_{2}}^{2}}{\mu ^{2}}]-m_{t_{1}}^{4}\ln
[\frac{m_{t_{1}}^{2}}{\mu ^{2}}])}{64\pi ^{2}}.\notag}
\end{align}%
\begin{figure}[h]
\begin{center}
\includegraphics[scale=.7]{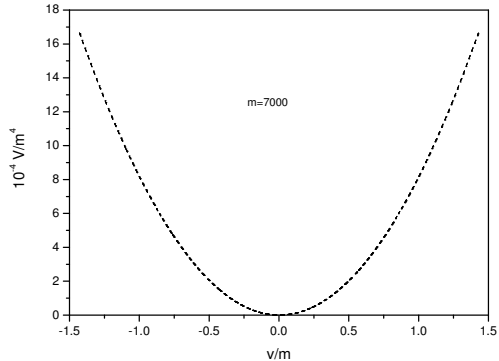}
\end{center}
\caption{The supersymmetric effective potential} \label{susyzero}
\end{figure}

Since relation (\ref{sp}) holds, the terms proportional to
$\frac{1}{L^{2}}$ cancel \cite {love}.\ Also, the minimum of the
potential vanishes at ${v=}$0 and supersymmetry is preserved.
Indeed, expanding (\ref{pot}) for small ${v}$ we get:
\begin{equation}
V\simeq m_{1}^{2}v^{2}+O(v^{3}).
\end{equation}%

In figure \ref{susyzero} we plot the effective potential for the
upper perturbative limit
$mL=1$. The other numerical values are chosen to be: ${m}_{1}{=}${200, }${m=}${7000, }${g}_{1}{%
=}${0.001, }${g=}${0.09, }${\mu =}${7000}.

\section{Addition of Explicit Supersymmetry Breaking Terms}

Let us now introduce into the Lagrangian (\ref{test}) various hard
supersymmetry breaking terms of the form $g_{3}\chi ^{2}\varphi
_{i}^{2}$ and $g_2\overline{\psi}\psi \chi$, where ${g}_{3}$ and
$g_2$ are dimensionless couplings (and recall $\chi =v+\varphi
_{u_{1}}/\sqrt{2}$).\ This terms, being non holomorphic and hard,
break supersymmetry explicitly (also these terms have moebiosity
zero).

Indeed the addition of such terms re-introduces quadratic
divergences in the theory, namely,
\begin{equation}
\Delta m_{scalar}=\frac{1}{8\pi^2}(l_s-l_f^2)\Lambda^2_{UV},
\end{equation}
with $\Lambda^2_{UV}$ a relevant upper cut-off of the theory and
$l_s$, $l_f$ boson and fermion couplings.

Since $\chi $ develops a vev, the fields coupled to it will
acquire an additional mass of the form $g_{3}v^{2}$ and $g_2v$. We
can add various combinations of the allowed terms. There exist two
class of phenomena occurring, depending on the supersymmetry
breaking terms we use.

The first type and, from a phenomenological point of view more
interesting, resembles the first order phase transition picture in
thermal field theory. Actually, regardless that there exist hard
supersymmetry breaking terms, supersymmetry is unbroken for small
values of the length of the compact dimension $L$. The minimum of
the potential is zero at $v=0$, $V(0)=0$. As the length of the
compact dimension increases, a second non supersymmetric minimum
is created after a critical length is reached. Then phenomena may
occur that can be described by the theory of metastable vacua (the
reader may find useful the papers \cite{alles,kirsten} where
similar issues are discussed in $S^1\times R^3$ but for a non
supersymmetric $\phi^4$ theory. Again non trivial topology induces
similar behavior of the effective potential).

The second type of phenomena describes a theory that supersymmetry
is unbroken when the length of the compact dimension is large and
as the radius decreases, supersymmetry breaks spontaneously after
a critical length. Thus supersymmetry is broken only for small
lengths. Although this is rather curious for a four dimensional
model, we shall present it because it may have application to
extra dimensional models.

\subsection{A term of the form $g_{3}\chi ^{2}\varphi _{u_{2}}^{2}$}

\begin{figure}[h]
\begin{center}
\includegraphics[scale=.7]{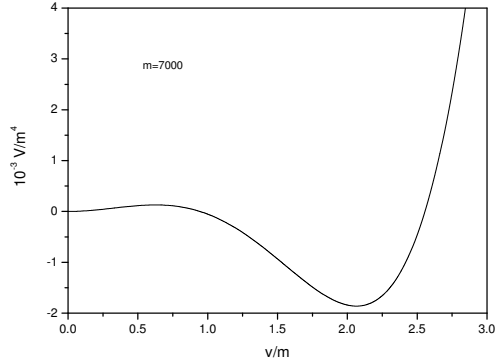}
\end{center}
\caption{The continuum effective potential}
\label{continuumnonsusy}
\end{figure}

\begin{figure}[h]
\begin{center}
\includegraphics[scale=.8]{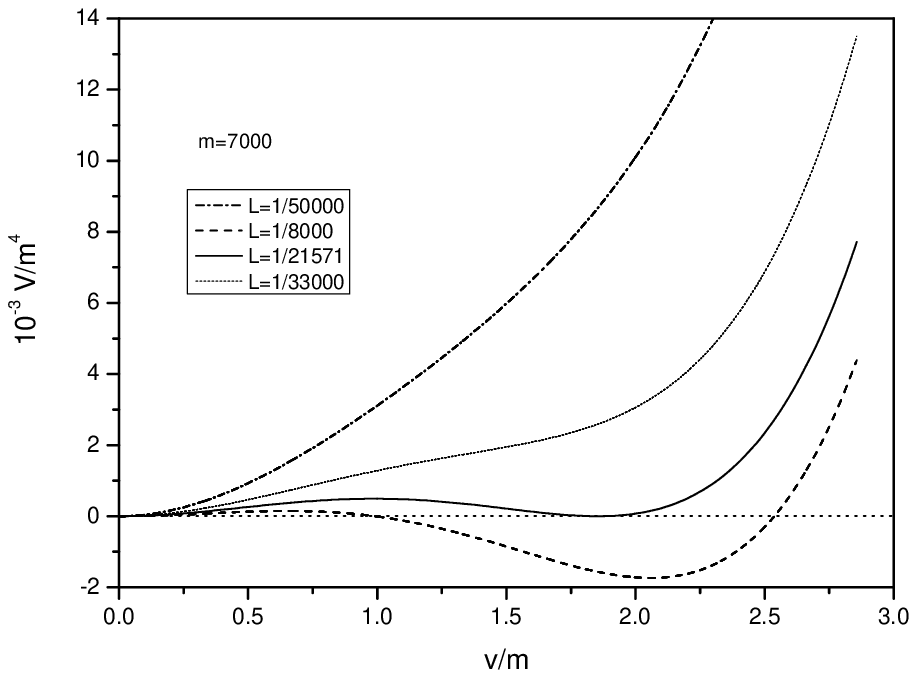}
\end{center}
\caption{The effective potential including finite size
corrections} \label{nonsusy}
\end{figure}

We introduce into the Lagrangian (\ref{test}) an interaction term
between the two untwisted scalar fields, of the form $-g_{3}\chi
^{2}\varphi _{u_{2}}^{2}$. Since $\chi =v+\varphi
_{u_{1}}/\sqrt{2}$, the scalar field $\varphi _{u_{2}}$ will
acquire an additional mass term of the form $g_{3}v^{2}$. This
way, the masses of the fields now become:
\begin{eqnarray}
m_{b_{1}}^{2} &=&m_{1}^{2}+3g_{1}m_{1}v+3g_{1}^{2}v^{2}/2 \\
\ m_{b_{2}}^{2}
&=&m_{1}^{2}+g_{1}m_{1}v+g_{1}^{2}v^{2}/2+g_{3}v^{2}  \notag
\\
m_{t_{1}}^{2}\bigskip &=&m^{2}+4gmv+4g^{2}v^{2}+g^{2}m_{1}v/\sqrt{2}%
+g^{2}g_{1}v^{2}/4  \notag \\
\ m_{t_{2}}^{2}
&=&m^{2}+4gmv-g^{2}m_{1}v/\sqrt{2}-g^{2}g_{1}v^{2}/4  \notag
\\
m_{f_{1}} &=&m_{1}+g_{1}v,\ m_{f_{2}}=m+2gv.  \notag
\end{eqnarray}
As expected, supersymmetry is now broken and relation (\ref{str})
becomes,
\begin{equation}
2m_{f_{1}}^{2}-m_{b_{1}}^{2}-m_{b_{2}}^{2}=g_{3}v^{2},\
m_{t_{1}}^{2}+m_{t_{2}}^{2}=2m_{f_{2}}^{2}.
\end{equation}%
One can see that the supersymmetric minimum at $v=0$ is still
preserved. Indeed, $V$ can be
written as:%
\begin{equation}
V\simeq (m_{1}^{2}+\frac{g_{3}}{768\pi ^{2}L^{2}})v^{2}+O(v^{3}).
\end{equation}%
We can see that in the continuum limit (infinite $L$), the
supersymmetric vacuum becomes metastable and a second non
supersymmetric vacuum appears. Including finite size corrections,
we see that for small ${L}$ the effective potential has a unique
supersymmetric minimum at $v=0$. As ${L}$ increases, a second
minimum develops, which becomes supersymmetric at the critical
value ${L}_{c}{=}\frac{1}{21571}$.\ When ${L}\!>\!{L}_{c}$ the
second minimum is non supersymmetric and becomes energetically
more preferable than the supersymmetric one
\cite{linde,zeldovich}.
This said behavior of the potential is valid whenever $g_{3}\gg g_{1}$ and for $%
\frac{m_{1}}{m}\ll g_{3}$. Using the same numerical values as
before, we plot the effective potential for ${g}_{3}{=}${0.5},
first in the continuum limit (figure~\ref{continuumnonsusy}), and
then including ${L}$ dependent corrections (figure~\ref{nonsusy}).

Let us discuss the above results. $g_{3}$, $g_{1}$ are couplings
among the untwisted superfields, $g_{3}$ corresponding to the
supersymmetry breaking term. If the $g_{3}$ interaction is
stronger than $g_{1}$ and if the mass (${m}$) of the twisted
superfield is larger than the untwisted one (${m_{1}}$), then the
following phenomenon occurs. For small length ${L}$ of the compact
dimension, supersymmetry is not broken (figure~\ref{nonsusy}). As
${L}$ grows larger, a second minimum appears which is not
supersymmetric (${L}\!>\!{L}_{c}$). There exists a small barrier
separating the two minima (figure~\ref{nonsusy}), and there exists
the possibility of quantum barrier penetration between them. This
resembles the first order phase transition picture of thermal
field theories.

Upon closer examination, we can see that in the continuum limit,
the supersymmetric vacuum becomes metastable and a second non
supersymmetric vacuum appears. Including finite size corrections,
we see that for small ${L}$ the effective potential has a unique
supersymmetric minimum at $v=$0. As ${L}$ increases, a second
minimum develops, which becomes supersymmetric at the critical
value ${L}_{c}{=}\frac{1}{21571}$.\ When ${L}\!>\!{L}_{c}$ the
second minimum breaks supersymmetry and becomes energetically more
preferable than the supersymmetric one \cite{linde,zeldovich}.
This said behavior of the potential is always valid whenever $g_{3}\gg g_{1}$ and for $%
\frac{m_{1}}{m}\ll g_{3}$. Using the same numerical values as
before, we plot the effective potential for ${g}_{3}{=}${0.5},
first in the continuum limit (figure~\ref{continuumnonsusy}), and
then including ${L}$ dependent corrections (figure~\ref{nonsusy}).

\subsection{A term of the form $g_{3}\chi ^{2}\varphi _{t_{1}}^{2}$}

Let us try something different now. We add an interaction among a
twisted boson and the untwisted boson that acquires vev, namely
$-g_{3}\chi ^{2}\varphi _{t_{1}}^{2}$. Since $\chi =v+\varphi
_{u_{1}}/\sqrt{2}$ the twisted boson $\varphi _{t_{1}}$ will have
additional contribution to it's tree order mass. The masses now
read,
\begin{eqnarray}
m_{b_{1}}^{2} &=&m_{1}^{2}+3g_{1}m_{1}v+3g_{1}^{2}v^{2}/2 \\
\ m_{b_{2}}^{2} &=&m_{1}^{2}+g_{1}m_{1}v+g_{1}^{2}v^{2}/2 \notag
\\
m_{t_{1}}^{2}\bigskip &=&m^{2}+4gmv+4g^{2}v^{2}+g^{2}m_{1}v/\sqrt{2}%
+g^{2}g_{1}v^{2}/4+g_3v^2  \notag \\
\ m_{t_{2}}^{2}
&=&m^{2}+4gmv-g^{2}m_{1}v/\sqrt{2}-g^{2}g_{1}v^{2}/4  \notag
\\
m_{f_{1}} &=&m_{1}+g_{1}v,\ m_{f_{2}}=m+2gv.  \notag
\end{eqnarray}
As expected $m_{t_{1}}^{2}+m_{t_{2}}^{2}-2m_{f_{2}}^{2}\neq 0$,
since supersymmetry is hard broken.

An interesting phenomenon occurs for this term and for a class of
other terms as we shall see. In detail, when the length of the
compact dimension is small, supersymmetry is broken and becomes
restored when the radius increases (and overcomes a critical
length $L_c$).

\begin{figure}[h]
\begin{center}
\includegraphics[scale=1.0]{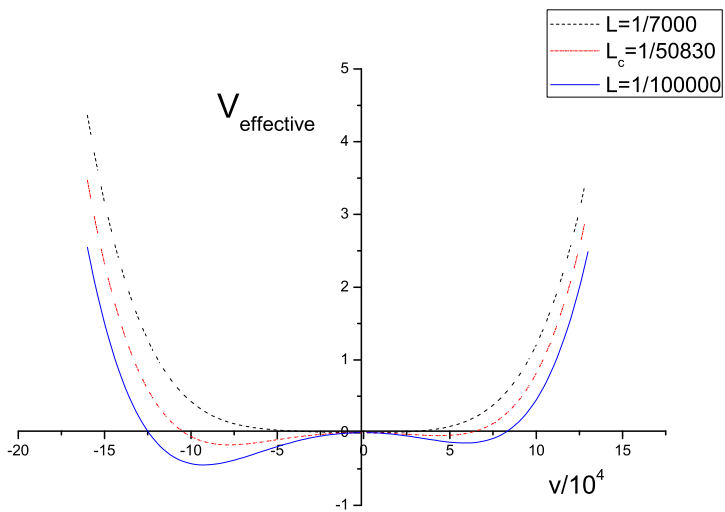}
\end{center}
\caption{Inverse Supersymmetry Breaking} \label{inverse1}
\end{figure}

This is strange and rather counterintuitive to what would be
expected from a phenomenologically correct four dimensional
theory. However we describe it since it might be useful to extra
dimensional physics. Also, as we shall see in the next section,
the whole behavior resembles the inverse symmetry breaking of
continuous symmetries at finite temperature. Let us call it
''inverse supersymmetry breaking'' for brevity. This said behavior
can appear when $g_3$ is of the order of $\frac{m_1}{m}$ or for
values smaller, that is $g_3\leq \frac{m_1}{m}$ when only the term
$g_{3}\chi ^{2}\varphi _{t_{1}}^{2}$ appears in the Lagrangian
(recall that in the previous subsection, the metastable vacua
phenomena of the previous subsection occurred when
$\frac{m_{1}}{m}\ll g_{3}$). This whole phenomenon is well seen in
figure~\ref{inverse1}. We used the following numerical values, ${m}_{1}{=}${200, }${m=}${7000, }${g}_{1}{%
=}${0.001, }${g=}${0.09, }${\mu =}${7000} and $g_3=0.05$.
As can be seen in figure~\ref{inverse1} the phenomenon looks like
a second order phase transition with the length of the compact
dimension playing the role of the temperature. No barrier appears
between the vacua at $v=0$ and at $v\neq 0$. The study was limited
to perturbation preserving values of $L$. As we see for large $L$
($L=1/7000$) supersymmetry is unbroken and start's to break at
$L_c=1/50830$. As the length decreases, the breaking is more
profound. The two non supersymmetric vacua are not equivalent.

\begin{center}\label{table1}
\begin{tabular}{|c|c|}
  \hline
  $g_3$ & $L_c^{-1}$ \\
  \hline
  0.1 & 32319 \\
  \hline
  0.07 & 41352 \\
  \hline
  0.05 & 50839 \\
  \hline
  0.03 & 67994 \\
  \hline
  0.01 & 121950 \\
   \hline
  0.007 & 145900 \\
   \hline
  0.005 & 173083 \\
   \hline
  0.003 & 223990 \\
   \hline
  0.001 & 388990 \\
   \hline
  0.0005 & 557000 \\
   \hline
  0.0001 & 1232000 \\
  \hline
  0.00005 & 1740000 \\
   \hline
  0.00001 & 3942000 \\
\hline
\end{tabular}
\\ \medskip{ \bfseries{Values of} $g_3$ \bfseries{and corresponding} $L_c$ }
\end{center}

\begin{figure}[h]
\begin{center}
\includegraphics[scale=.9]{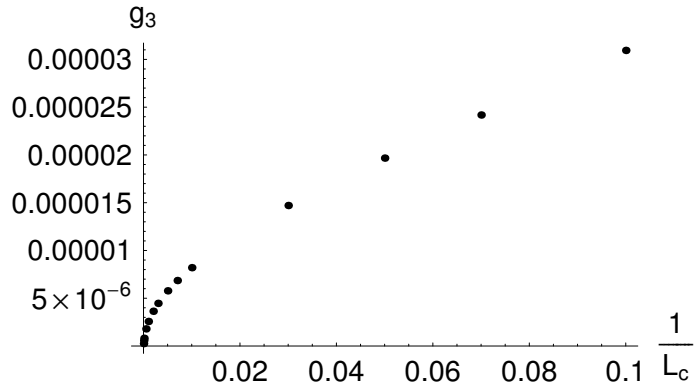}
\end{center}
\caption{Plot of $g_3$ and $L_c$} \label{plotl3lc}
\end{figure}

\begin{figure}[h]
\begin{center}
\includegraphics[scale=.9]{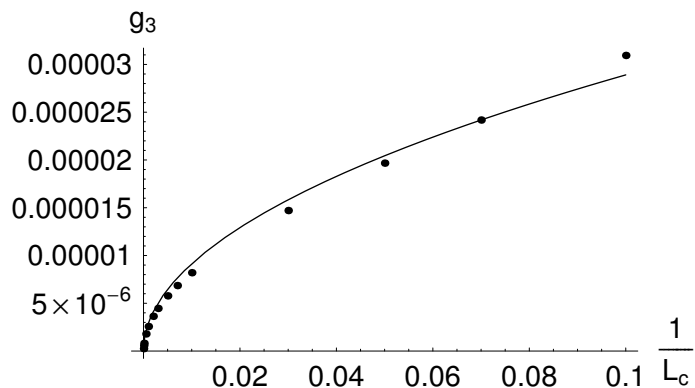}
\end{center}
\caption{Fit of the curve $g_3-L_c$}\label{fitl3lc}
\end{figure}

We tried to find how $L_c$ changes under a change of $g_3$. In the
table we present the values of $g_3$ and the corresponding values
of $L_c$, and in figure \ref{plotl3lc} we plot the dependence. In
figure \ref{fitl3lc} we fit the curve with a continuous function.
The dots are the values that appear in the table, while the
continuous line corresponds to the function $0.000091\sqrt{x}$.
Thus the dependence of $g_3$ as a function of $L_c$ is roughly,
\begin{equation}
g_3\sim 0.000091\sqrt{L_c}.
\end{equation}

\subsection{Other Hard Supersymmetry Breaking terms and Discussion}

We shall not present here the detailed study of all the allowed
terms (this will be done in \cite{oikonomouunderpreparation}). We
shall discuss only the main results.

A term of the form $g_2\overline{\psi_i}\psi_i\chi$ always breaks
supersymmetry and non of the previous phenomena occurs. As we
shall see in the next section this is similar to symmetry non
restoration where a continuous symmetry is broken and never get's
restored.

The combined addition of terms
$-\frac{1}{2}g_2\overline{\psi_t}\psi_t\chi-g_3\chi ^2\varphi
_{t_1} ^2$ (that is interaction of the untwisted scalar $\chi$
with a twisted fermion and twisted boson) causes ''inverse
supersymmetry breaking'' as previous. The conditions that must
hold in order this occurs are the same as before ($g_3\leq
\frac{m_1}{m}$) and in addition $g_2 \ll g_3$. For this condition
the behavior of supersymmetry breaking is well described from
figure \ref{inverse1}. Also for $g_2=0.0001$ the $g_3-L_c$
dependence is similar to that of figure \ref{fitl3lc}.
Particularly the $g_3-L_c$ dependence in this case, is described
by,
\begin{equation}
g_3\sim 0.000091\sqrt{L_c},
\end{equation}
which is the same as before. This dependence is the same for all
the cases we studied. Thus this motivates us to think that there
is a universal behavior of $g_3$ as a function of $L_c$.

Of course all effects we described in this section appear in one
loop level and within perturbative limits. Thus there exist the
danger that all these effects are an artifact of perturbation
theory. However the theory is at some level supersymmetric and one
loop corrections may be adequate enough \cite{martin}. A detailed
study should include higher loop corrections. In the next section
similar problems-considerations are encountered and discussed.

\section{Continuous and Discrete Symmetry Inverse Symmetry Breaking, Restoration and Non-restoration at Finite Temperature}

In this section we review some conceptually similar phenomena to
the above. The difference is that symmetries are studied at finite
temperature and the symmetry is not supersymmetry but a continuous
global $O(N_1)\times O(N_2)$ or a discrete $Z_2\times Z_2$. As we
shall see symmetry non-restoration and inverse symmetry breaking
phenomena occur naturally when similar terms to
$\frac{1}{2}g_2\overline{\psi_t}\psi_t\chi$, $g_3\chi ^2\varphi
_{t_1} ^2$ appear in the Lagrangian.

Symmetry non-restoration means that a symmetry broken at $T=0$
never gets restored at high temperatures. Inverse symmetry
breaking means that an unbroken symmetry at $T=0$ may be
spontaneously broken at high temperature.

These phenomena occur in field theories when cross interactions
are included among the scalar fields similar to the bosonic hard
supersymmetry breaking term $g_{3}\chi ^{2}\varphi _{u_{2}}^{2}$.
Similar to this term scalar interactions and also Yukawa terms
like the ones of the previous sections are frequently used in the
theory of reheating after inflation. Actually this similarities
motivated us to use such terms in order to see what their effect
would be on supersymmetry breaking. We shall discuss these in the
end of this section.

First let us describe the inverse symmetry breaking phenomenon.
Consider a theory with real scalar fields $\phi_1$ and $\phi_2$
described by the $O(N_1)\times O(N_2)$ globally symmetric
Lagrangian,
\begin{align}\label{onlagrang}
&L=\frac{1}{2}\partial _{\mu }\phi_1\partial _{\mu }\phi_1
+\frac{1}{2}\partial _{\mu }\phi_2\partial _{\mu }\phi_2
+\frac{1}{2}m_1^2\phi_1^2+\frac{1}{2}m_2^2\phi_2^2+\frac{1}{4!}\lambda_1\phi_1^4
+\frac{1}{4!}\lambda_1\phi_2^4+\frac{1}{4!}\lambda \phi_1^2
\phi_2^4,
\end{align}
where $\phi_1$ and $\phi_2$ be real scalars with $N_1$ and $N_2$
components. In the above Lagrangian one of the global $O(N_i)$
symmetries may break at high temperature if the $\lambda$ coupling
takes negative values. Thus one of the two scalar fields $\phi_1$
or $\phi_2$ may acquire a non zero vacuum expectation value. Thus
at high temperature and for certain values of the parameters, the
initial $O(N_1)\times O(N_2)$ is broken to $O(N_1)$. This was
called inverse symmetry breaking and was first point out by
Weinberg \cite{weinbergantirestor} and extensively studied by many
authors \cite{bimonte1,bimonte2,pinto1,pinto2,dvali,pinto3,barut}.

In the case $N_1=1$ and $N_2=1$ the initial $O(N_1)\times O(N_2)$
symmetry reduces to a $Z_2\times Z_2$ symmetry and similar
arguments hold, in connection to the breaking of one of the two
discrete symmetries (for details see \cite{pinto1})

Symmetry non-restoration was used in \cite{dvali} to solve the
monopole problem in the $SU(5)$ GUT. Monopoles are usually
produced during phase transitions at temperatures of the order
$\sim10^{14}$GeV. As was proposed in \cite{dvali}, the symmetric
phase of $SU(5)$ was never realized, no matter how high the
temperature becomes. In that paper the interaction term $\alpha
\vert \chi_{45}\vert ^2 \vert H_{24}\vert$ appearing in the scalar
Kibble-Higgs sector is responsible for the non-restoration of the
$SU(5)$ symmetry. Actually the scalar interaction of $H_{24}$ and
$\chi_{45}$ gives negative contributions to the thermal masses and
one of those becomes negative. Once this happens the corresponding
Higgs field maintains a vacuum expectation value for high
temperatures and the symmetry is never restored. This phenomenon
occurs for certain values of the parameters (see
\cite{dvali,bimonte2}). However these results are very sensitive
and are altered when someone includes two loop calculations
\cite{bimonte2}. In the same spirit there are arguments based on
large N calculations which can show that symmetry non-restoration
is an artifact of perturbation theory. For an interesting
discussion on this see \cite{bimonte2} and references therein.

A cross term of the form $g\chi^2\phi^2$ is used in non
relativistic models in condensed matter physics, for example in
the coupled two field Bose gases. However the effect of this does
not break any of the initial symmetry patterns \cite{pinto3}.

In conclusion the intuitive approach to all phenomena at finite
temperature consists of the statement that symmetries broken at
small temperatures become restored at high temperature (in the
same class belong finite volume theories). Many field theory
models exist that belong to this class. Counter intuitive
phenomena occur in field theories with rich scalar sector.
Especially if the multi-scalar sectors interact weakly with
negative couplings then the phenomena known as inverse symmetry
breaking or symmetry non-restoration occur. This happens at high
temperature and refers to the spontaneous breaking of a symmetry
at high temperature. Usually the symmetry is a global \linebreak
$O(N_1)\times O(N_2)$ or for the case of symmetry non-restoration
a continuous, like $SU(5)$.

Although inverse symmetry breaking is counterintuitive, nature has
provided us with examples that systems are more symmetric at low
temperatures than at high temperature. For example the Rochelle
salt which at low temperatures is orthorhombic and after a
critical temperature becomes monoclinic. Similar phenomena occur
in liquid crystals (for more examples see \cite{barut} and
references therein).

\subsection{Reheating After Inflation and Thermal Inflation}

The process of reheating after inflation is one of the most
important features of the new inflationary scenario
\cite{lindebook}. The process of reheating is necessary in order
that the inflating vacuum like state of the universe transforms to
a hot Friedmann universe state. During the reheating process a
massive scalar field gives it's vacuum energy to lighter fermions
and bosons. The Lagrangian governing this process is,
\begin{align}\label{reheatinglagra}
&L=\frac{1}{2}(\partial_{\mu}\varphi)^2-\frac{m_{\varphi}}{2}\varphi^2+\frac{1}{2}(\partial_{\mu}\chi)^2-\frac{m_{\chi}}{2}\chi^2
+\overline{\psi }(i\gamma ^{\mu }\partial
_{\mu}-m_{\psi})\psi-\lambda \varphi
\overline{\psi}\psi-\frac{1}{2}g^2\varphi ^2\chi ^2.
\end{align}
The role of the inflaton fields is played from the scalar field
$\varphi$. The inflaton field decays to the particles $\psi$ and
$\chi$ due to the interaction terms $\lambda \varphi
\overline{\psi}\psi$ and $\frac{1}{2}g^2\varphi ^2\chi ^2$. Note
that we used similar terms in order to break supersymmetry hard
and all the effects we seen in the previous section are due to
these interaction terms. Also, in order reheating takes place, the
condition $m_{\varphi}\gg m_{\psi}, m_{\varphi}$ must hold
(remember that similar conditions hold in the supersymmetric model
we studied previously. There $m_1$ was the tree order mass of the
untwisted scalar field and $m$ the tree order mass of the twisted
scalar. One of the conditions we used is that the untwisted sector
has greater mass than the twisted sector, namely $m_1\gg m$. Also
the untwisted fermion has mass m).

So with the interactions $\lambda \varphi \overline{\psi}\psi$ and
$\frac{1}{2}g^2\varphi ^2\chi ^2$ the initially concentrated
energy (during inflation) to the field $\varphi$ is transferred to
particle creation through the processes $\varphi \rightarrow \psi
\psi $ and $\varphi \rightarrow \chi \chi$, and the universe
thermalizes \cite{reheatingpapers}.

The cross interaction terms of the form $\frac{1}{2}g^2\varphi
^2\chi ^2$ are used in some versions of hybrid inflation.

Let us briefly mention another application of cross interactions
between scalar fields. These terms are used to the thermal
inflation scenario. This is a modified version of the new
inflation and old inflation scenario
\cite{lindebook,thermalinflation}. In the thermal inflation
scenario, a Lagrangian that contains a term $\frac{1}{2}g^2\varphi
^2\chi ^2$ is used. This term is necessary in order a phase
transition takes place. Actually there is bump in the effective
potential that is solely created from this interaction term and
the phase transition is strongly first order. The universe
supercools in the false vacuum and after a critical temperature
tunnels to the true vacuum through bubble nucleation. At this
point thermal inflation ends (for details see
\cite{thermalinflation}).

Before closing this section we conclude that cross terms between
scalar fields and Yukawa interactions are frequently used in
particle models and these terms make the effective potential
locally unstable thus triggering first order phase transitions,
reheating and other processes. Also at high temperatures and for
specific values of the parameters, these terms may cause inverse
symmetry breaking or symmetry non restoration.

Our study involved these terms but the phenomena studied where
related to supersymmetry breaking and inverse supersymmetry
breaking when the space has a compact dimension. So the
resemblance between the two setups is quite clear. We shall
discuss on this later on.

\section{A Toy Cosmological Application}

In this section a brief qualitative application (although
fictitious) of one of the above results is presented. Consider a
toy universe that has just come out of it's strong gravity period
and it's particle content (matter) is described by (\ref{test})
with the addition of the hard supersymmetry breaking term
$g_{3}\chi^{2}\varphi _{u_{2}}^{2}$. The back-reaction of gravity
on field theory is considered small ({\it{i.e.}}{\,}field theory
calculations made in the previous part considering flat
background, are consistent and thus the metric fluctuations are
negligible). This toy universe's expansion is described by:
\begin{equation}
\mathrm{d}s^{2}=\mathrm{d}t^{2}-a^{2}(t)\mathrm{d}x_{1}^{2}-b^{2}(t)\mathrm{d}x_{2}^{2}-c^{2}(t)\mathrm{d}x_{3}^{2}
\label{metric},
\end{equation}
a homogeneous Clifford-Klein metric (known as Bianchi I
cosmological model), with ${x_1}$, ${x_2}$, ${x_3}$, as in
(\ref{flat}). In (\ref{metric}), $a(t)$, $b(t)$, $c(t)$, describe
the scale factors of the two infinite and of the compact dimension
respectively. Also we assume,
\begin{equation}
a(t)=b(t)=k{\,}c(t) \label{cond},
\end{equation}
with $k\gg{1}$.

Figure \ref{nonsusy} motivates us to think as follows: At small
lengths of the compact dimension the toy universe's ground state
is the supersymmetric vacuum although we had broken supersymmetry
using a hard term, something usually unexpected. As the
circumference of the compact dimension grows, the toy universe
''acknowledges'' the presence of the other true vacuum (in terms
of it's quantum one loop effective potential) and at some point
(bubbles of the new vacuum create within the false vacuum) quantum
penetrates to the other vacuum, the non supersymmetric one.
Therefore, at small lengths of the compact dimension,
supersymmetry was unbroken and as the radius grows, supersymmetry
breaks. It seems that using a compact dimension in the present
model, supersymmetry breaks dynamically after some critical radius
of the compact dimension, although supersymmetry is expected to be
broken for all lengths (this would exactly be the effect of a hard
supersymmetry breaking term).

Let us now do some toy cosmology on this toy universe. $V(0)$ is
the minimum of the effective potential at the origin (note
$V(0)=0$), and $V(v_1)$ the minimum after quantum barrier
penetration (the non supersymmetric vacuum). We assume this toy
universe has a cosmological constant which is chosen to be
$(8{\pi}G)^{-1}\Lambda=-\Delta{V}=-(V(v_1)-V(0))$ (a choice which
shall be explained below). Note that $\Lambda>0$.

The Friedmann equation describing it's evolution is:
\begin{equation}
{\Big{(}}\frac{\dot{a}}{a}{\Big{)}}^2
=\frac{8{\pi}G}{3}{\Big{(}}\rho+\frac{\Lambda}{8{\pi}G}{\Big{)}},
\end{equation}
referred to the $x_1$ dimension (we omit the analysis on the other
dimension and to the compact one since it is similar (\ref{cond}).
For details see \cite{gongcharov}).

In the early post quantum gravity period, this toy universe is at
the $V(0)$ vacuum state, the energy density is $\rho={V(0)}=0$.
The Friedmann equation reads:
\begin{equation}
{\Big{(}}\frac{\dot{a}}{a}{\Big{)}}^2
=\frac{8{\pi}G}{3}{\Big{(}}\frac{\Lambda}{8{\pi}G}{\Big{)}},
\end{equation}
and without getting into much detail (see \cite{gongcharov}), an
inflationary solution (corresponding to a flat universe) follows
in all space dimensions, being of the form,
\begin{equation}
a(t)\sim e^{\sqrt{\Lambda}t},
\end{equation}
with $a(t)=b(t)=k{\,}c(t)$. Note that the rate of the expansion is
the same for all dimensions. During the inflation period of this
toy universe, its quantum vacuum state is the supersymmetric
vacuum (false vacuum), until for some length quantum tunnelling
occurs (due to one loop quantum effects), and the new vacuum state
is $V(v_1)$, the new minimum of the effective action. During the
quantum vacuum penetration, the energy release (something like
latent heat) \cite{vilenkin} is of the order $L^{-4}_p$ which
thermalizes the matter content at a temperature $T_p$, with
\begin{equation}
L^{-4}_p\sim T^{4}_p,
\end{equation}
$L_p$ and $T_p$ characterizing the ''phase transition'' point.

After thermalization, the energy density is $\rho \sim
T^{4}_p+V(v_1)$ and the Friedmann equation reads:
\begin{equation}
{\Big{(}}\frac{\dot{a}}{a}{\Big{)}}^2 \sim
\frac{8{\pi}G}{3}{\Big{(}}T^{4}_p{\Big{)}},
\end{equation}
(we fixed $\Lambda$ in order to cancel the value of $V(v_1)$). So
after vacuum penetration the toy universe follows a radiation
dominated expansion (note that the maximum temperature ever
reached was the thermalization temperature $T_p$ \cite{vilenkin}).

Note that the above picture has many similarities with the
strongly supercooled first order phase transitions of the early
universe (old inflationary scenario). Let us point out its main
features. Start with a toy universe filled with fermions and
bosons interacting in a non supersymmetric way (due to explicit
hard breaking). The toy universe is at a supersymmetric vacuum
(unexpectedly) when it's magnitude (specifically the compact
dimension magnitude) is small, but as it evolves spatially
(inflation in our setup) quantum penetrates to a non
supersymmetric vacuum, which is energetically preferable. So at
the early toy universe's epoch, supersymmetry was not broken (at
least the vacuum quantum state did not realize broken susy),
although the matter content of it, interacts in a non
supersymmetric way, but supersymmetry dynamically breaks (through
quantum tunnelling) \cite{linde,zeldovich} when the toy universe
evolves at larger sizes.

Let us note here that in order this toy universe is realistic, one
must deal with defects (monopoles, domain walls) and with the
cosmological experimental observations that do not suggest non
trivial topology in the spatial dimensions. Maybe domain walls may
be avoided in first order phase transitions but if we want to
include GUTs in this universe we can not avoid defects (only if
the temperature after the quantum penetration is smaller compared
to the temperature that the defects are created, then defects
maybe be avoided). Even if one deals with defects, the non trivial
topology problem remains, so the magnitude of the compact
dimension must be larger than the particle horizon (which can be
achieved during inflation).

\section{Conclusions}

In this paper we studied a simple supersymmetric model in
$S^1\times R^3$ spacetime topology. We discussed how topology can
affect the boundary conditions of the fields and we seen that in
$S^1\times R^3$ bosons and fermions can have periodic and
anti-periodic boundary conditions along the compact dimension.
Also we discussed how supersymmetry is broken spontaneously and
how this can be avoided in terms of the boundary conditions that
the fields obey. We confirmed these by calculating the effective
potential of the theory.

Next we introduced in the Lagrangian interaction terms among
scalars and fermions. These terms break supersymmetry hard in
$R^4$ topology. However two class of phenomena occur in $S^1\times
R^3$:
\begin{itemize}
\item When an interaction among the two untwisted scalars is
added, a term of the form $g_{3}\chi ^{2}\varphi _{u_{2}}^{2}$,
supersymmetry remains unbroken for small values of the radius of
the compact dimension, and as the length increases, breaks after a
critical length. This occurs when $\frac{m_{1}}{m}\ll g_{3}$. This
phenomenon resembles first order phase transitions of finite
temperature field theories. Also we applied this to a toy
cosmological model, which described the evolution of a universe
with an initial cosmological constant and filled with the
aforementioned fields.

\item The addition of an interaction among the scalar of the
untwisted sector that develops a vev and the twisted scalars or
the twisted fermions results in a very peculiar phenomenon.
Particularly when certain conditions hold (similar to the
aforementioned) supersymmetry is broken for small lengths and as
the radius increases, becomes restored after a critical length.
This resembles conceptually second order phase transitions.
\end{itemize}
The last is similar to inverse symmetry breaking phenomena at
finite temperature, that appear in field theories with rich scalar
sector. In that case a symmetry unbroken at low temperatures may
break at high temperature. In our case at small lengths
supersymmetry breaks while remains unbroken for large radius
values. We called this inverse supersymmetry breaking for brevity.
The terms in the Lagrangian that trigger inverse symmetry breaking
are the same that trigger inverse supersymmetry breaking. Also
these terms appear in the theory of reheating and in the thermal
inflation. In the study we realized that there is a universality
in the $g_3-L_c$ dependence. We shall present these in detail in
\cite{oikonomouunderpreparation}.

Finally let us discuss the physical significance of ''inverse
supersymmetry breaking''. This phenomenon is not so appealing to a
four dimensional theory. What would be expected is that
supersymmetry should be unbroken for small values of the radius of
the compact dimension and breaks dynamically at large distances.
What happens here is the converse. For large values of the radius,
supersymmetry is unbroken and breaks dynamically for small values
of the radius. However this would be interesting for a five
dimensional model. Imagine a theory where supersymmetry is
unbroken for large radius of the compact dimension  and breaks
dynamically for small values of the compact dimension. It is an
interesting task to find what this mechanism (which is basically
coupling interplay between interactions) has to say for the radius
stabilization mechanism of extra dimensional models.

\end{document}